\documentclass[reprint,twocolumn,secnumarabic,amssymb, aps, prl]{revtex4-2}
\usepackage{booktabs} 

\usepackage{graphicx}
\usepackage{gensymb}
\usepackage{amsmath}
\usepackage{braket}
\usepackage{lipsum}

\usepackage[colorlinks=true,linkcolor=black,citecolor=black,urlcolor=black]{hyperref}

\usepackage{wrapfig}

\newcounter{appendixsection}

\newcounter{appendixsubsection}

\usepackage{tikz, ifthen}
\usepackage{tikz-cd}
\usetikzlibrary{arrows}
\usetikzlibrary{angles,quotes} 
\usetikzlibrary{snakes}
\usetikzlibrary{intersections}
\usetikzlibrary{shapes.geometric}
\usetikzlibrary{decorations.pathmorphing, patterns,shapes}
\usetikzlibrary{decorations.markings}
\usepackage{xparse}
\ExplSyntaxOn
  \seq_new:N \g_my_seq
  \seq_set_from_clist:Nn \g_my_seq {crimson,dodgerblue,forest}

  \cs_new:Npn \GetItem #1
    { \seq_item:Nn \g_my_seq {#1} }
\ExplSyntaxOff

\tikzset{
	partial ellipse/.style args={#1:#2:#3}{
		insert path={+ (#1:#3) arc (#1:#2:#3)}
	}
}

\tikzset{
	mid arrow/.style={postaction={decorate,decoration={
				markings,
				mark=at position .575 with {\arrow[#1]{stealth}}
	}}},
	near arrow/.style={postaction={decorate,decoration={
				markings,
				mark=at position .275 with {\arrow[#1]{stealth}}
	}}},
	far arrow/.style={postaction={decorate,decoration={
				markings,
				mark=at position .800 with {\arrow[#1]{stealth}}
	}}},
}

\pgfdeclarepatternformonly{south west lines}{\pgfqpoint{-0pt}{-0pt}}{\pgfqpoint{3pt}{3pt}}{\pgfqpoint{3pt}{3pt}}{
	\pgfsetlinewidth{0.4pt}
	\pgfpathmoveto{\pgfqpoint{0pt}{0pt}}
	\pgfpathlineto{\pgfqpoint{3pt}{3pt}}
	\pgfpathmoveto{\pgfqpoint{2.8pt}{-.2pt}}
	\pgfpathlineto{\pgfqpoint{3.2pt}{.2pt}}
	\pgfpathmoveto{\pgfqpoint{-.2pt}{2.8pt}}
	\pgfpathlineto{\pgfqpoint{.2pt}{3.2pt}}
	\pgfusepath{stroke}}

\begin{document}
\title{
High-fidelity entangling gates and nonlocal circuits with neutral atoms}

\author{
Simon~J.~Evered$^{1,*}$, 
Muqing~Xu$^{1,*}$,
Sophie~H.~Li$^{1,*}$,
Alexandra~A.~Geim$^{1}$,
J.~Pablo~Bonilla~Ataides$^{1}$,
Marcin~Kalinowski$^{1}$,
Dolev~Bluvstein$^{1}$,
Nishad~Maskara$^{2}$,
Christian~Kokail$^{3,1,4}$,
Markus~Greiner$^{1}$,
Vladan~Vuleti\'c$^{5}$,
Mikhail~D.~Lukin$^{1}$
}

\affiliation{
$^1$Department~of~Physics,~Harvard~University,~Cambridge,~MA~02138,~USA\\
$^2$Massachusetts~Institute~of~Technology,~Center~for~Theoretical Physics,~Cambridge,~MA~02138,~USA\\
$^3$QuEra Computing Inc., Boston, MA 02135, USA\\
$^4$ITAMP, Harvard-Smithsonian Center for Astrophysics, Cambridge, MA, 02138, USA\\
$^5$Department~of~Physics~and~Research~Laboratory~of~Electronics,~Massachusetts~Institute~of~Technology,~Cambridge,~MA~02139,~USA\\
$*$ These authors contributed equally to this work.
}

\begin{abstract}
Creation and manipulation of entanglement with low error is essential in quantum information systems. In practice, two-qubit entangling gates constitute a dominant error source, limiting circuit depths and performance in fault-tolerant architectures. Using a neutral-atom quantum processor, we realize entangling CZ gates with a high Rabi frequency smooth-amplitude pulse, employing state-selective readout and qubit reuse for fast calibration, and achieve state-of-the-art fidelities of 99.854(4)\% which improve to 99.941(3)\% upon loss postselection, with stable performance for 10 hours. We then use these low-error gates in quantum circuits with coherent atom rearrangement. We first benchmark performance by creating and disentangling cluster states, and subsequently implement scrambling circuits featuring longer-range connectivity to study non-locally entangled states generated through chaotic dynamics. These results pave the way towards deep-circuit, efficient fault-tolerant quantum computation.
\end{abstract}
\maketitle

High-fidelity entanglement is at the heart of quantum science and technology, enabling wide-ranging applications in computing, simulation, metrology, and sensing~\cite{Preskill2018}. For instance, error rates around  $0.1\%$ are generally believed to be necessary to enable efficient performance of quantum error correction (QEC), central to realizing utility-scale fault-tolerant quantum computation~\cite{Fowler2012,Beverland2023,Gidney2025,Gu2026}. In particular, achieving such high-fidelity performance for nonlocal quantum circuits over large numbers of qubits could enable the implementation of efficient high-rate QEC codes~\cite{Bravyi2024, Xu2024} as well as significant algorithmic speedups with transversal gate architectures~\cite{Cain2024, Zhou2024}.

Neutral atom processors have recently emerged as a leading quantum computing approach, combining nonlocal qubit connectivity~\cite{Bluvstein2022}, scalable control of large qubit numbers~\cite{Manetsch2025,Tao2024}, continuous operation~\cite{Chiu2025,Li2025,Norcia2024}, and the ability to efficiently implement quantum error correction~\cite{Bluvstein2023,Rodriguez2025,Muniz2025_reuse,Reichardt2025}. In this approach, atomic qubits are entangled using controlled-phase gates based on Rydberg blockade~\cite{Jaksch2000,Saffman2010}. The fidelity of these entangling operations has rapidly advanced~\cite{Levine2019,Graham2019,Evered2023, Ma2023, Peper2025,Muniz2025, Radnaev2025}, with recent benchmarking demonstrations reaching $\sim 99.7\%$ fidelity~\cite{Tsai2025}. Moreover, methods for loss detection and erasure conversion have been shown in a range of settings, including loss-postselected fidelities reaching $\sim 99.8-99.85\%$~\cite{Senoo2025,Lin2026,Lib2026} and improved quantum error correction with erasure information~\cite{Zhang2025,Rines2025} including below-threshold performance~\cite{Bluvstein2025}.

In this article, we demonstrate a new state-of-the-art in entangling CZ gate fidelity for neutral atom qubits, with a raw fidelity of 99.854(4)\% that improves to 99.941(3)\% upon postselection on neither atom being lost. Our approach combines high Rabi frequency gates with a smooth-amplitude pulse profile based on optimal control~\cite{Jandura2022,Pagano2022,Evered2023} with fast calibration based on loss-resolved qubit readout~\cite{Bluvstein2025}, and we demonstrate how this fidelity can be maintained for over 10 hours without recalibration. We utilize these gates to realize quantum circuits based on coherent atom motion, first benchmarking similarly high-fidelity CZ gate performance in the making and unmaking of cluster states. Then, we design and implement nonlocal circuits featuring longer-range atom transport, which exhibit super-ballistic scrambling of quantum information. We study the nature of these nonlocal circuits by sampling from the output distribution, and show consistency with numerical simulations of the chaotic dynamics. These advances open the door for efficient realization of quantum algorithms with nonlocal connectivity as well as deep-circuit fault tolerant quantum computation with high-rate QEC codes. \\

\begin{figure*}
\centering
\includegraphics[width=1\textwidth]{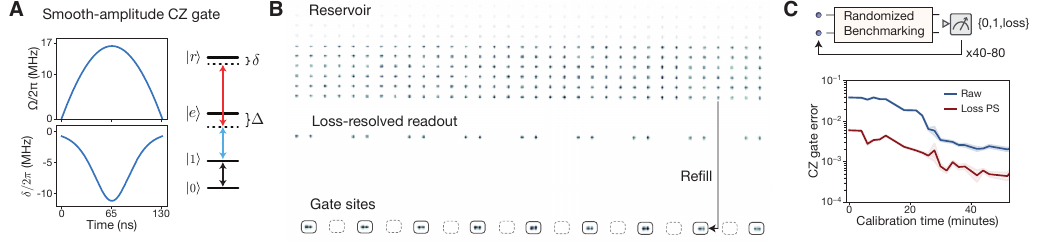}
\caption{\textbf{CZ entangling gates in a neutral atom processor with fast cycle rates.} (\textbf{A}) Rydberg gate realized through a smooth-amplitude gate pulse at roughly $\Omega = 2\pi\times17$ MHz peak Rabi frequency, with a detuning profile $\delta(t)$. Two-photon excitation scheme consists of transitions at 420 nm and 1015 nm, exciting from qubit state $\ket{1}$ to the Rydberg state $\ket{r} = 53S_{1/2}, m_J = -1/2$ through the intermediate state manifold $\ket{e} = 6P_{3/2}$ with intermediate state detuning $\Delta = 2\pi\times7.8$ GHz (see fig.~\ref{fig:ED_ExperimentalSystem}B for more detailed excitation diagram). (\textbf{B}) Images of the atoms separated into processor zones, with gate sites used for benchmarking and high-fidelity quantum circuits, and state-selective readout to perform loss detection. Fast calibration of CZ gates is enabled by qubit reuse and refilling lost atoms from a large atom reservoir. (\textbf{C})  Randomized benchmarking is performed over 40-80 repetitions, with a 20-30 Hz cycle rate. Example calibration of the CZ gate from starting parameters near the theoretical prediction, showing both raw and loss-postselected CZ error. Each point corresponds to the fitted maximum of a parabola scan over one of the gate parameters, with error bars given by the shaded region.}
\label{fig1}
\end{figure*}

\begin{center}
\textbf{Neutral atom entangling gates}
\end{center}

Our approach makes use of a reconfigurable neutral atom processor architecture~\cite{Bluvstein2023,Bluvstein2025}, with several key upgrades to achieve and explore low error regimes. Quantum information is encoded in long-lived hyperfine clock states of $^{87}$Rb atoms, and entangling gates are performed by exciting the pair of atomic qubits to $n=53$ Rydberg states~\cite{Evered2023}. We use a gate profile based on optimal control~\cite{Jandura2022, Pagano2022} with a smooth amplitude and phase profile, as shown in Fig.~\ref{fig1}A~\cite{Evered2023}, and additional amplitude corrections to compensate for pulse imperfections~\cite{Ma2023,supplementary_materials}. This gate is optimized to suppress scattering from the intermediate excited state $\ket{e}$ by maximizing population in the dark state not containing $\ket{e}$ ~\cite{Evered2023}. To suppress errors from Rydberg state lifetime, we utilize high Rabi frequencies reaching a peak of roughly $2\pi\times 17$ MHz. The required high laser intensities induce substantial lightshifts on the Rydberg transition and necessitate careful stabilization of Rydberg laser positions and powers to minimize the detrimental effect on ground-Rydberg coherence $T_2^*$~\cite{supplementary_materials}. Additionally, the coupling to the other Rydberg level $\ket{r'}$ with $m_J = +1/2$ grows as Rabi frequency is increased, which we address by applying a stronger magnetic field (larger Zeeman splitting) to suppress off-resonant excitations (fig.~\ref{fig:ED_ExperimentalSystem}B).

Our quantum processor utilizes up to 442 atomic qubits arranged into separated zones (Fig.~\ref{fig1}B), with 17 gate sites used for high-fidelity benchmarking and quantum circuits, and a large reservoir for refilling lost atoms. We employ state-selective readout using spin-to-position conversion~\cite{Bluvstein2025}, which enables us to directly detect atom loss errors at the end of the circuit. By reusing qubits and refilling lost atoms from the reservoir, we are able to perform fast calibration of the CZ gate, for up to 80 repetition cycles within a single round of atom loading, at a cycle rate of 20-30 Hz (Fig.~\ref{fig1}C). We find that calibrating the gate from scratch, which consists of scanning each gate parameter in series, typically takes $<40$ minutes, with calibrated gate parameters remaining stable such that subsequent recalibration is faster. \\

\begin{figure*}
\centering
\includegraphics[width=1\textwidth]{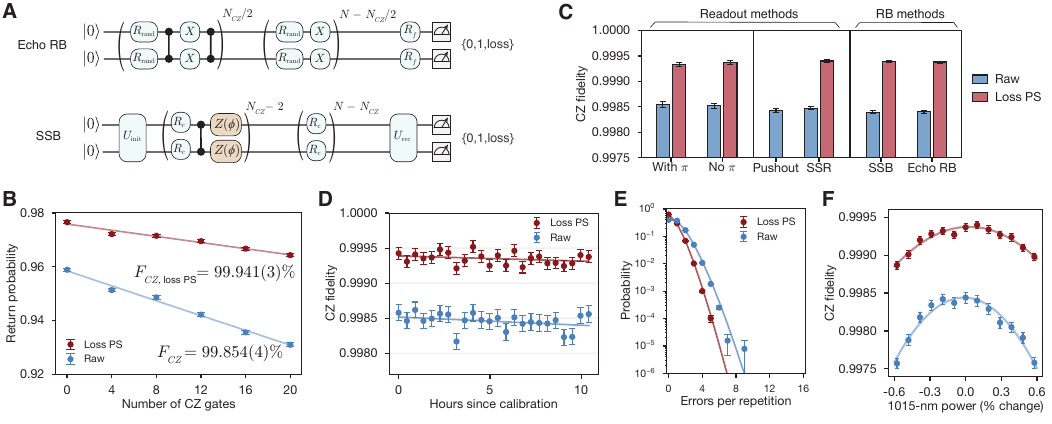}
\caption{\textbf{Stable low-error CZ gate performance.} (\textbf{A}) Benchmarking methods based on repeated CZ gate applications: global echo randomized benchmarking (``echo RB")~\cite{Evered2023} and symmetric stabilizer benchmarking (``SSB")~\cite{Tsai2025}. (\textbf{B}) Return probability as a function of number of CZ gates, fitted to an exponential decay, giving a raw fidelity of 99.854(4)\% and a loss-postselected fidelity of 99.941(3)\%. (\textbf{C}) Comparison of CZ gate fidelity measured with different benchmarking protocols: echo RB with and without an extra single-qubit $\pi$ pulse before readout, resonant pushout of $\ket{1}$ versus state-selective readout (SSR), and SSB versus echo RB. (\textbf{D}) Stability of the CZ fidelity measured for over 10 hours with echo RB, with no recalibration of the gate. Each data point consists of a fit to the decay of return probabilities for 0, 8, 12, and 20 CZ gates. (\textbf{E}) Number of errors per repetition (raw and loss-postselected), where a repetition consists of benchmarking on two rows of 16 atoms brought into the entangling zone, for the data shown in \textbf{D}, consisting of 127,680 repetitions. Solid curves are sums of individual Poissonian fits for each number of CZ gates. (\textbf{F}) Sensitivity of the raw and loss-postselected CZ gate fidelity under perturbation in the 1015-nm Rydberg beam power, keeping the gate pulse parameters unchanged.}
\label{fig2}
\end{figure*}

\begin{center}
\textbf{Benchmarking low-error CZ gates}
\end{center}

To characterize the resulting CZ gate performance, we utilize randomized benchmarking (RB) based on repeated application of CZ gates interspersed with single-qubit rotations (Fig.~\ref{fig2}A). First, we use a global echo RB sequence using a single-qubit $\pi$ pulse in between pairs of CZ gates~\cite{Evered2023}, and measure a raw fidelity of 99.854(4)\% fidelity and a fidelity of 99.941(3)\% when postselecting on neither atom being lost at the end of the sequence (Fig.~\ref{fig2}B). Our detection of these loss errors can be used to improve QEC thresholds~\cite{Baranes2026, Bluvstein2025} and quantum simulation results~\cite{Evered2025}. In particular, we find that the CZ error due to atom loss is 0.087(5)\%, comprising $\sim 60\%$ of the total error. We measure a loss probability of 0.054(2)\% per atom per CZ gate, which exceeds half of this total loss error and indicates correlated loss of both atoms during the gates; this information can be used in QEC decoding~\cite{Perrin2026}. Additionally, we explore the effects of leakage to other $m_F$ levels~\cite{Tsai2025} in the ground state $5S_{1/2},$ $F=1,2$ manifolds (from Rydberg $T_1$ and intermediate state scattering) through measurements of $F=1$ populations at the end of the RB sequence, which increase as 0.008(1)\% per atom per CZ gate (fig.~\ref{fig:ED_Leakage})~\cite{supplementary_materials}.

To test the robustness of the gate performance, we measure the CZ fidelity across several different benchmarking and readout methods (Fig.~\ref{fig2}C). First, we compare echo RB to the symmetric stabilizer benchmarking (SSB) protocol developed in Ref.~\cite{Tsai2025}, which is theoretically insensitive to single-qubit gate errors and requires calibration of the single-qubit phase from the CZ gate (fig.~\ref{fig:ED_SSB}). We find that both the raw and loss-postselected fidelities are equivalent for the two RB methods, consistent with no measurable effect from single-qubit gate errors or miscalibration of the single-qubit phase of the gate. We additionally compare state-selective readout with destructive spin-to-loss readout based on resonant blowout of one of the qubit states, and again find comparable results for echo RB (fig.~\ref{fig:ED_Benchmarking}B). Finally, we measure similar fidelities with and without an extra single-qubit $\pi$ pulse before readout, which suggests that any residual effects of leakage on the RB results are relatively balanced between the $F=1$ and $F=2$ levels (fig.~\ref{fig:ED_Benchmarking}A).

To test the stability of our gates, we track the fidelity over long durations (Fig.~\ref{fig2}D). Remarkably, we observe stable performance at $\sim 99.85\%$ fidelity for over 10 hours without any recalibration of the gate. For this large dataset, consisting of roughly 2 million runs of the RB sequence averaged over the array, we further observe that both the raw and loss-postselected return probabilities follow Poisson distributions with no observed large-scale correlated error events spanning the system (Fig.~\ref{fig2}E). We find that the residual drift in the gate fidelity typically originates from drifts in the Rydberg laser powers and positions, such that peak performance can generally be recovered by small adjustments in beam powers and positions, with calibrated gate pulse parameters remaining unchanged. To understand the magnitude of the effect of drifts, in Fig.~\ref{fig2}F we intentionally adjust the power of 1015-nm Rydberg beam away from optimal, and observe that the gate is sensitive to $\sim0.1\%$-scale power changes. We further find that the sensitivity of the CZ gate to perturbations in the Rydberg beam powers and positions is well-captured by numerical error simulations that include the effects on gate detuning and $T_2^*$ (fig.~\ref{fig:ED_Drifts}). These observations suggest that neutral-atom devices with sufficiently stable Rydberg beams can achieve high-fidelity performance over long durations passively, and over even longer timescales with automated stabilization of beam positions and powers, for example via real-time steering of QEC performance~\cite{Sivak2025}.

\begin{figure}
\centering
\includegraphics[width=1\columnwidth]{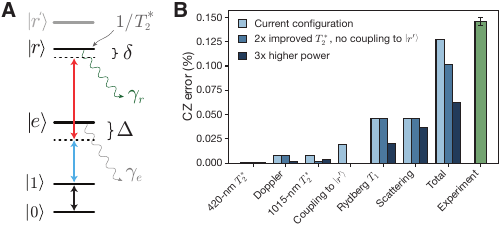}
\caption{\textbf{Error breakdown and path to further improvement.} (\textbf{A}) Illustration of error sources for Rydberg entangling gates, including scattering from the intermediate state $\gamma_e$, Rydberg state lifetime $\gamma_r$, ground-Rydberg coherence $T_2^*$, and coupling to the other Rydberg state $\ket{r'} = 53S_{1/2}, m_J = +1/2$. (\textbf{B}) Numerical error budget showing contributions of individual errors, total theoretical errors, and experimental data. Error breakdowns are shown for the current configuration, as well as projected values attainable with technical improvements and higher laser intensities~\cite{supplementary_materials}.
}
\label{fig3}
\end{figure}

The remaining CZ gate errors can be largely explained by a numerical error model, with the dominant contributions being decay from the Rydberg state due to finite lifetime and from the intermediate excited state in the form of off-resonant scattering (Fig.~\ref{fig3}). Using our numerical error model, we project that raw CZ gate fidelities at the level of 99.9-99.95\% could be enabled by improved $T_2^*$, suppressed coupling to $|r'\rangle$, and higher laser intensities (fig.~\ref{ed_fig_fidelity_projections}). \\

\begin{center}
\textbf{High-fidelity circuits with atom motion}
\end{center}

\begin{figure}
\centering
\includegraphics[width=1\columnwidth]{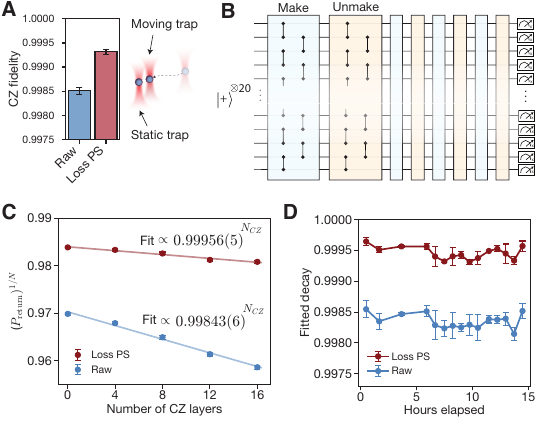}
\caption{\textbf{High-fidelity quantum circuits with atom motion.} (\textbf{A}) Raw and loss-postselected CZ fidelity measured with echo RB, with one atom in a static trap generated by a spatial light modulator (SLM) and one atom in a moving trap generated by an acousto-optic deflector (AOD). (\textbf{B}) Protocol for making and unmaking one-dimensional cluster states to study CZ gate performance in a quantum circuit. (\textbf{C}) Decay of global return probability $P_{\text{return}}$ as a function of the number of total CZ gates, fit to an exponential decay, for $N=20$ qubits. (\textbf{D}) Stability of the fitted decay over time, without recalibration.}
\label{fig4}
\end{figure}

We next extend our benchmarking of CZ gates to quantum circuits based on coherent rearrangement of atomic qubits. Several additional challenges can arise when incorporating high-fidelity entangling gates into quantum circuits. In our randomized benchmarking circuits, we place the two atoms in two traps generated by the same spatial light modulator (SLM), which allows us to put the atoms uniformly close together ($\approx1.7\mu$m spacing). By contrast, in quantum circuits, we perform gates between atoms in stationary SLM traps and atoms in moving traps generated by acousto-optic deflectors (AODs)~\cite{Bluvstein2023}. We find that to perform these gates with high fidelity, it is important for the spacing between traps to be uniform across the array, and for the focus of the two trap arrays to be in the same plane~\cite{supplementary_materials}. When properly calibrated, we can realize similarly high CZ gate performance, as shown in Fig.~\ref{fig4}A.

To understand the CZ performance in quantum circuits with atom motion, we perform a characterization circuit corresponding to repeatedly making and disentangling 20-atom cluster states (Fig.~\ref{fig4}B). We measure the probability that all atoms in the array return to the initial state $\ket{0}^{\otimes 20}$ after 0, 4, 8, 12, and 16 nearest-neighbor CZ gate layers, defined as the global return probability $P_{\text{return}}$. By fitting $P_{\text{return}}$ as a function of the total number of CZ gates applied across the entire qubit array, we infer the corresponding CZ gate fidelity (Fig.~\ref{fig4}C). We find a raw fidelity of 99.843(6)\% and a loss post-selected value of 99.956(5)\%, similar to the fidelity measured in gate benchmarking sequences. While not a rigorous benchmarking method, we observe in numerical simulations that the extracted fidelity from this characterization circuit approximately agrees with the CZ gate fidelity (fig.~\ref{fig:ED_mirror_fidelity}). To test the stability of the CZ gate performance within this circuit, we leave the experiment running for 15 hours, and find that the extracted CZ fidelity remains high without any recalibration (Fig.~\ref{fig4}D). \\

\begin{figure*}
\centering
\includegraphics[width=0.8\textwidth]{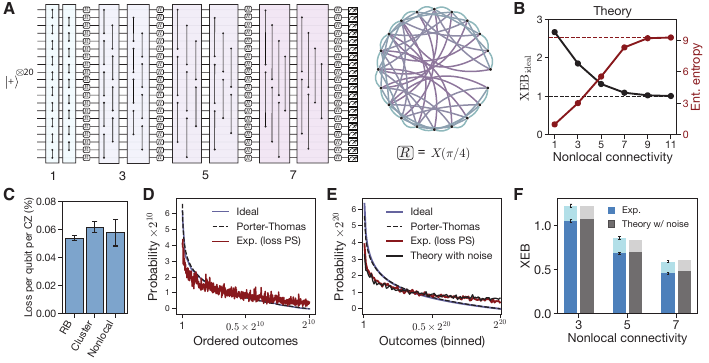}
\caption{\textbf{High-fidelity scrambling circuits with nonlocal connectivity.} (\textbf{A}) Nonlocal quantum circuit implementation. CZ gates of increasing nonlocal connectivity (with labels given below the circuit) are interspersed with single-qubit gates, $R = X(\pi/4)$. The resulting connectivity for the 20-qubit circuit is illustrated on the right. (\textbf{B}) Theoretical values for the half-system-size von Neumann entanglement entropy and $\text{XEB}_{\text{ideal}}$ as a function of the degree of nonlocal CZ gates added to the circuit (``nonlocal connectivity"). Dashed lines indicate the Page entropy value and $\text{XEB}_{\text{ideal}}=1$. (\textbf{C}) Loss per qubit per CZ gate compared for different circuits in this work: randomized benchmarking (RB), the cluster state circuit in Fig.~\ref{fig4}B, and the nonlocal quantum circuit in Fig.~\ref{fig5}A. (\textbf{D}) Sampled outcome distribution for the 10-qubit nonlocal quantum circuit, run with the depth shown in \textbf{A}, comparing the experimental data to the ideal theory expectation and a Porter-Thomas distribution. (\textbf{E}) Sampled outcome distribution for the 20-qubit nonlocal circuit, with the outcomes binned in 100 groups by their probability. Comparison to numerical simulations with noise is shown. (\textbf{F}) Experimentally measured XEB for the 20-qubit circuit, for different degrees of nonlocal connectivity, compared to numerical simulations with noise. Solid bars are raw data, and lighter bars correspond to postselection on no atom loss.}
\label{fig5}
\end{figure*}

\begin{center}
\textbf{Chaotic nonlocal circuits}
\end{center}

Finally, we explore high-fidelity entanglement in more complex circuits with longer-range qubit connectivity. Owing to the high degree of connectivity, such circuits are known to be effective in scrambling quantum information, and can be difficult to simulate classically~\cite{Ghosh2024,Decross2025}. In Fig.~\ref{fig5}, as an example, we design and study a class of nonlocal circuits with increasingly long-range gates as a function of depth, interspersed with $\pi/4$ single-qubit gates between layers. This circuit structure leads to super-ballistic scrambling of quantum information, in which entanglement entropy saturates at the Page entropy, the near-maximal value expected for a random pure state, in a depth which scales as $\sqrt{N}$ for system size $N$ (Fig.~\ref{fig5}B and fig.~\ref{fig:ED_nonlocal_theory}). Our circuits can also be interpreted as nonlocal kicked Ising models, where the corresponding nearest-neighbor circuit does not scramble (fig.~\ref{fig:ED_nonlocal_theory}A)~\cite{supplementary_materials}. The output random states are related to the physics of black holes~\cite{Hayden2007}, and have been used for benchmarking quantum devices~\cite{Cross2019,Choi2023} and quantum advantage experiments~\cite{Arute2019}. 

We first assess fidelity of the CZ gate in these nonlocal quantum circuits applied on 20 qubits through measurement of atom loss, which we find to be a useful proxy for the in-situ performance of the gate in circuits. We measure the loss per qubit per CZ gate to be consistent across the randomized benchmarking, cluster state, and nonlocal circuits throughout this work (Fig.~\ref{fig5}C). Additionally, we observe that the loss per shot for both circuits follows a Poisson distribution, with no evidence of large-scale loss events (fig.~\ref{fig:ED_CircuitImplementation}C).

Next, we sample from the output of this circuit, by measuring all qubits in the Pauli $Z$ basis and plotting the probability of each outcome ordered by theoretical expectation. 
For random states, the ideal probability distribution $p(x)$ is an exponentially distributed Porter-Thomas distribution PT$(x)$~\cite{Porter1956}. 
Such probability distributions can be studied using cross-entropy benchmarking (XEB), where $\text{XEB} = 2^n\sum_x q(x)p(x) - 1$ for outcomes $x$ and system size $n$. The ideal value in the absence of noise is given by the second moment of the probability distribution, $\text{XEB}_{\text{ideal}} = 2^n\sum_x p^2(x) - 1$, often called the collision probability, for which the Porter-Thomas distribution PT$(x)$ has $\text{XEB}_{\text{ideal}} = 1$.
We find that the theoretical output distributions of our circuits converge to PT$(x)$ for sufficient depth (Fig.~\ref{fig5}B and fig.~\ref{fig:ED_nonlocal_theory}).

Experimentally, we start by studying the distribution in a 10-qubit circuit, where we can measure the probability of all $2^N$ outcomes (fig.~\ref{fig:ED_CircuitImplementation}B). We find that the resulting distribution is similar to PT$(x)$ for outcomes $x$ (Fig.~\ref{fig5}D). Next, we extend to sampling from the 20-qubit circuit by binning the outcomes into groups (Fig.~\ref{fig5}E). We find that our results are well described by numerical simulations with noise, with decoherence making the distribution closer to the uniform distribution. To characterize the $2^{20}$ sampled outcomes more quantitatively, in Fig.~\ref{fig5}F we compare XEB for the 20-qubit circuit sampling distribution to numerical simulations, and find good agreement for the same error model across three different circuit depths~\cite{supplementary_materials}. While we do not directly probe the CZ gate fidelity as in Fig.~\ref{fig4}, the combination of low atom loss per gate and agreement with numerical simulations suggests high-fidelity performance in these complex nonlocal circuits. \\

\begin{center}
\textbf{Discussion and outlook}
\end{center}

Our results pave the way for practical fault-tolerant quantum computation with atomic processors. Corresponding to more than 3$\times$ improvement from our previous CZ gates used in below-threshold performance of a surface code logical qubit~\cite{Bluvstein2025}, these higher fidelities, in combination with improving single-qubit operations, should enable $>10\times$ below-threshold QEC performance, while leveraging the ability to detect atom loss errors~\cite{Baranes2026}. These improved fidelities were achieved with $\approx1.7\times$ higher Rabi frequency, requiring suppression of coupling to $\ket{r'}$, stronger blockade with closer spacing, improved Rydberg coherence ($T_2^*$), as well as fast calibration of the gate pulse~\cite{supplementary_materials}. We find that the remaining CZ gate errors are dominated by Rydberg state lifetime and intermediate state scattering. Through technical improvements to further suppress coupling to $\ket{r'}$ and improve the Rydberg $T_2^*$, we expect that reaching 99.9-99.95\% raw CZ fidelity will be achievable by scaling up the available laser power (fig.~\ref{ed_fig_fidelity_projections}).

Future work will focus on implementing these high-fidelity entangling gates across larger atomic systems with up to $ 10,000$ or more qubits, for instance leveraging efficient beam shaping~\cite{Ebadi2021}, fast beam scanning~\cite{Rines2025, Wei2026, Bytyqi2026}, higher-power and pulsed lasers~\cite{Chew2022}, and different Rydberg excitation schemes \cite{Ma2023,Tsai2025}. Furthermore, our techniques can be extended to other gate schemes, including entangling gates at constant atomic velocity~\cite{Lib2026} and multi-qubit gates such as CCZ~\cite{Evered2023,Cao2024}.

The high-fidelity quantum circuits demonstrated in this work form the basis for wide-ranging explorations of highly-entangled quantum systems and fault-tolerant architectures. Extensions of our nonlocal circuits could be used to simulate nonlocal quantum models, including efficient encoding of fermions with qubits~\cite{Maskara2025} and models for gravity~\cite{Sahay2025}. Additionally, such circuits can be used to probe complex properties of quantum many-body systems, providing optimal system-size-independent sample complexity for measuring linear observables~\cite{Huang2020}. Most significantly, implementing complex nonlocal circuits with high fidelity should enable the realization of high-rate quantum error correcting codes to achieve low error rates on large numbers of logical qubits~\cite{Bravyi2024,Xu2024}. \\

\noindent\textbf{Acknowledgments:} We thank Oana Bazavan, Matteo Bergonzoni, Oriol Rubies Bigorda, Sergio Cantu, Michael Gullans, Dominik Hangleiter, Jonas Helsen, Simon Hollerith, Dhruv Kedar, Sylvain de Léséleuc, Alexander Lukin, James MacArthur, Tom Manovitz, Ognjen Marković, Takuya Matsubara, Stefan Ostermann, Guido Pupillo, Xiangkai Sun, and Tout Wang for fruitful discussions. We especially thank Simon Hollerith for his careful reading and feedback on the manuscript.
\textbf{Funding:}
We acknowledge financial support from the US Department of Energy (DOE/LBNL Quantum Systems Accelerator Center, grant number DE-AC02-05CH11231), the IARPA and the Army Research Office, under the Entangled Logical Qubits program (Cooperative Agreement Number W911NF-23-2-0219), DARPA MeasQuIT program (grant number HR0011-24-9-0359), the Center for Ultracold Atoms (an NSF Physics Frontier Center, grant number PHY-2317134), the National Science Foundation (grant numbers PHY-2012023, CCF-2313084, and NVQL grant PHY-2410716), Wellcome Leap Quantum for Bio program, and QuEra Computing. S.J.E. acknowledges support from the National Defense Science and Engineering Graduate (NDSEG) fellowship. M.X. acknowledges support from the Harvard Quantum Initiative Postdoctoral Fellowship in Science and Engineering. \\

\noindent\textbf{Author contributions:} 
S.J.E., M.X., S.H.L., and A.A.G. contributed to building the experimental setup, performed measurements, analyzed the data, and developed gate error understanding. J.P.B.A., N.M., and C.K. carried out theoretical analysis and numerical simulations of quantum circuits. M.K. and D.B. contributed to understanding error sources and initial development of gate improvements. All work was supervised by M.G., V.V., and M.D.L. All authors contributed to the project vision, discussed the results, and contributed to the manuscript. \\

\noindent\textbf{Competing interests:} 
M.G., V.V., and M.D.L. are co-founders and shareholders, V.V. is Chief Technology Officer, M.G. is a consultant, M.D.L. is Chief Scientist, J.P.B.A. has served as a consultant, N.M. was a part-time employee, and C.K. is an employee of QuEra Computing. \\ 

\noindent\textbf{Data, code, and materials availability:} All data are available in the manuscript and the supplementary materials.

\let\oldaddcontentsline\addcontentsline
\renewcommand{\addcontentsline}[3]{}
\bibliographystyle{sciencemag}
\bibliography{references.bib}

\let\addcontentsline\oldaddcontentsline

\clearpage
\onecolumngrid
\begin{center}
    \textbf{\Large Supplementary Materials}
\end{center}
\normalsize

\setcounter{equation}{0}
\setcounter{figure}{0}
\setcounter{table}{0}
\makeatletter
\renewcommand{\theequation}{S\arabic{equation}}
\renewcommand{\thefigure}{S\arabic{figure}}
\renewcommand{\thetable}{S\arabic{table}}
\setlength\tabcolsep{10pt}
\setcounter{secnumdepth}{2}
\renewcommand\thesection{\arabic{section}}
\newcommand\numberthis{\addtocounter{equation}{1}\tag{\theequation}}
\newcommand{\insertimage}[1]{\includegraphics[valign=c,width=0.04\columnwidth]{#1}}
\tableofcontents

\section{Experimental system \label{Sec:expt}}

\begin{figure}
\centering
\includegraphics[width=1\textwidth]{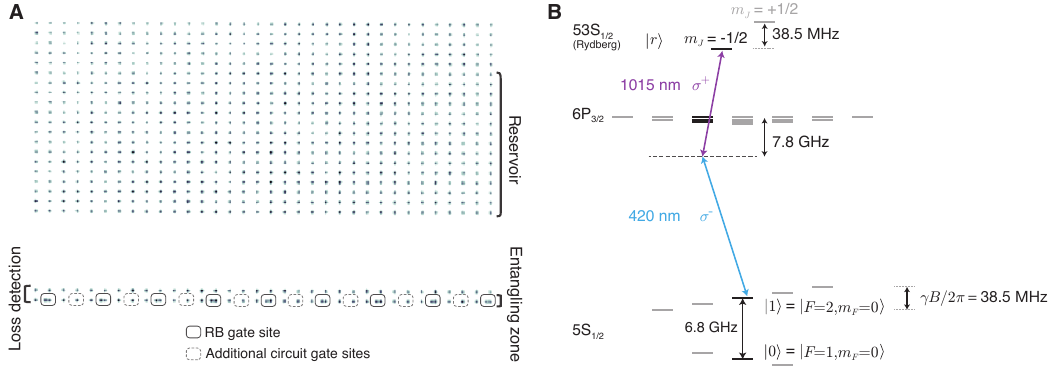}
\caption{\textbf{Architecture for high-fidelity gates and circuits.}
(\textbf{A}) Atom array layout utilized in this work, with a large reservoir zone for fast cycle rates, an entangling zone with gate sites demarcated (dark sites are the subset used for randomized benchmarking), and additional sites in the entangling zone used for state-selective readout~\cite{Bluvstein2025}. The top five rows of sites are used for initial loading and rearrangement of atoms into the reservoir. (\textbf{B}) Level diagram for Rydberg excitation to $n=53$ Rydberg state $\ket{r}$. The lasers used for other aspects of the architecture, including for Raman single-qubit gates and state-selective readout, are described in our previous work~\cite{Bluvstein2025}.
}
\label{fig:ED_ExperimentalSystem}
\end{figure}

Our experiments begin by stochastically loading rubidium-87 atoms from a magneto-optic trap (MOT) into an array of static 852-nm optical tweezers generated with a spatial light modulator (SLM, Hamamatsu X13138-02) using 795-nm lambda-enhanced gray molasses for enhanced loading of $\sim75\%$~\cite{Brown2019GrayMolasses}. We rearrange the atoms using a second set of 852-nm optical tweezers generated from two-dimensional acousto-optic deflectors (AODs, DTSX-400, AA Opto-Electronic). For benchmarking gates and performing high-fidelity quantum circuits, we use a reservoir of 374 atoms, and bring up to 20 atoms at a time into the entangling zone (Fig.~\ref{fig:ED_ExperimentalSystem}A). We image the atoms with a 0.65-NA objective (Special Optics) onto a CMOS camera (Hamamatsu ORCA-Quest C15550-20UP). Compared to our previous work benchmarking high-fidelity gates in Ref.~\cite{Evered2023}, here we only use 3D polarization gradient cooling (PGC), as we find that errors from atomic temperature form a small part of our overall error budget (Fig.~\ref{fig3}B).

The qubit states are encoded in the $m_F=0$ states in the ground hyperfine manifolds, $\ket{0} = \ket{F=1,m_F=0}$ and $\ket{1} = \ket{F=2,m_F=0}$, with coherence time $T_2>1$s~\cite{Bluvstein2025}.
At the start of the quantum circuits, the qubits are prepared in $\ket{0}$ using Raman-assisted optical pumping. A global Raman beam illuminating the entire array, including atoms in all zones, is used to implement single-qubit operations~\cite{Levine2021}, with an intermediate state detuning of $\sim550$ GHz to reduce scattering error and achieve fidelities for BB1 composite pulses of $\sim 99.99\%$~\cite{Bluvstein2025}.

We implement loss-resolved qubit readout, as described in Ref.~\cite{Bluvstein2025}. In this approach, state-selective qubit readout is achieved using a one-dimensional state-selective optical lattice potential, which pins one of the two qubit states and allows us to move the other with a moving AOD trap. We split the two qubit states into two different vertical positions in the entangling zone, and then we move them in parallel to the bottom of the reservoir zone to image them. This approach retains the majority of atoms, enabling us to reuse the qubits in subsequent repetitions, in addition to refilling from the large reservoir. Throughout this work, we perform 40-80 repeated experiments before loading an atom array from the MOT.

The entangling zone used for benchmarking CZ gates and realizing high-fidelity circuits consists of 17 total gate sites. For benchmarking, we typically use a sparse array of 8 gate sites, in order to reduce the effect of decay of Rydberg states to nearby $P$ states on the measured fidelity~\cite{Evered2023}. Since this effect does not impact quantum circuits, in which CZ gates are spaced further apart in time due to additional time overhead from movement~\cite{Bluvstein2025}, we use the denser arrangement of gate sites for the circuit implementations.

\subsection{Rydberg excitation}

For realizing CZ gates, we excite atoms from the qubit state $\ket{1}$ to the Rydberg state 53$S_{1/2}$, using a smooth-amplitude gate (see section below for details)~\cite{Evered2023}.  We use circularly polarized 420-nm and 1015-nm light in a two-photon transition, with an intermediate state detuning of $7.8$ GHz (Fig.~\ref{fig:ED_ExperimentalSystem}B). We operate with the pair of atoms within each gate site separated by about 1.73$\mu$m, where the blockade energy is around 950 MHz. We pulse the traps off for a few hundred nanoseconds during the gate.

The laser systems for Rydberg excitation are the same as reported in Ref.~\cite{Evered2023}. By operating red-detuned from the intermediate state, we ensure that the $\sigma^-$ polarized 420-nm light is further detuned for the other qubit state $\ket{0}$. The $\sigma^+$ polarized $1015$-nm laser primarily couples to $\ket{r}$ with $m_J = -1/2$ in this magnetic field configuration, but also couples to the other $m_J = +1/2$ Rydberg state $\ket{r'}$ with a two-photon Rabi frequency suppressed by a factor of three. To mitigate the effect of this spurious coupling, we increase our bias magnetic field strength to $B=13.7$ G, with a larger $38.5$ MHz splitting between the two Rydberg levels than in our previous work. As in previous work, staying close to the dark state during the two-photon Rydberg excitation is crucial for reducing scattering error~\cite{Evered2023}. This suppression is achieved by choosing the detuning of the Rydberg gate to start with a positive sign. As a result, the peak Rabi frequency occurs at negative detuning. The combination of magnetic field direction and laser polarization ensures that at peak Rabi frequency, we remain far detuned from $\ket{r'}$, to reduce unwanted excitation to this Rydberg state.

We use high Rabi frequencies reaching a peak of $\Omega \approx 2\pi\times 17$ MHz in this work. The effective single-photon Rabi frequencies of the two legs of the two-photon transition are $\Omega_{\text{420,peak}} = 2\pi\times 670$ MHz and  $\Omega_{\text{1015}} = 2\pi\times 460$ MHz for the peak 420-nm and 1015-nm intensities, respectively. The two-photon Rabi frequency is given by $\Omega = \sqrt{3/4}\frac{\Omega_{\text{420,peak}}\Omega_{\text{1015}}}{2\Delta}$, where the prefactor of $\sqrt{3/4}$ comes from the presence of multiple intermediate states~\cite{Evered2023}. To achieve these high Rabi frequencies, we generate circularly symmetric beams with Gaussian waists of roughly 30 microns and 35 microns for the 420-nm and 1015-nm beams, respectively, to cover a one-dimensional entangling zone. The $420$-nm laser beam profile is Gaussian and is created by displaying a phase grating on the SLM (Hamamatsu).
The $1015$-nm laser beam profile is a circular tophat profile created by the SLM (Hamamatsu). These Rydberg beams are stabilized on cameras between each sequence, using active feedback to piezo-controlled mirrors.

\subsection{CZ gate profile and calibration}

We use the following parametrization for the smooth-amplitude gate in this work, with phase profile $\phi(t)$ and Rabi frequency profile $\Omega_{420}(t)$:


\begin{align*}
     \Omega_{420}(t)/ \Omega_{420,\mathrm{peak}} &= \Omega_{0} + A\mathrm{sech}[\omega_{\Omega}\tau]^{\alpha},\\
     \Omega_{0} &= \frac{A_{\mathrm{min}}\cosh[\omega_{\Omega}T/2]^{\alpha}-A_{\mathrm{max}}}{\cosh[\omega_{\Omega}T/2]^{\alpha} - 1}, \\
     A  &= \frac{(A_{\mathrm{max}}-A_{\mathrm{min}})\cosh[\omega_{\Omega}T/2]^{\alpha}}{\cosh[\omega_{\Omega}T/2]^{\alpha} - 1}, \\
      \phi(t) &= \delta_0\,\tau+B\tanh(\lambda\, \tau),
\end{align*}

\noindent where $\tau = t-T/2$.

Theoretical parameters for ideal smooth amplitude gates are given in our earlier work~\cite{Evered2023}. The following parameters are the experimental values used in Fig.~\ref{fig2}B: 

\begin{align*}
    A_{\mathrm{min}} &= -0.1565,& B&= -2\pi\times 0.3569,\\
    A_{\mathrm{max}} &= 0.9622,& \lambda &= 1.1195\,\Omega,\\
    \omega_\Omega &=  -0.2962\,\Omega, & \delta_0 &=  0.775\,\Omega,\\
    \alpha &= 0.3,
\end{align*}

\noindent For the other experiments, the parameters are similar, with small adjustments from performing recalibration. To compensate for pulse imperfections arising from our experimental setup, we additionally apply amplitude pulse corrections in the form of Chebyshev polynomials $T_n(t) = \cos\!\big(n \arccos (t/t_{\text{max}})\big)$~\cite{Ma2023}, which we sum together with coefficients $c_n$ and multiply by the amplitude profile. We find that most of these corrections have optimal coefficients near zero ($<0.01$). There is, however, a finite value of $c_1 = -0.214$ for the linear correction $T_1(t) = t/t_{\text{max}}$. Adding this term improves the fidelity at the $\sim 0.02-0.03\%$ level. We attribute its nonzero optimal value to imperfections in our experimental setup which make the pulse asymmetric.

We calibrate the CZ gate experimentally by starting from close to a theoretically optimal gate, and then scanning individual gate parameters to maximize the return probability of the echo RB sequence. Compared to our previous work~\cite{Evered2023}, here we are able to detect loss errors using state-selective readout, and we observe that this can be helpful in gaining intuition for the calibration process. First, the loss postselected fidelity is very sensitive to certain parameters such as detuning, but relatively insensitive to other parameters such as pulse duration. This observation is consistent with the role of different gate parameters: certain parameters more strongly affect the gate trajectory, which determines the population returning to the ground hyperfine manifold and loss error, whereas other parameters primarily determine the accumulated phase during the gate and the resulting phase error. Relatedly, we observe that at the beginning of calibration, the optimal parameters for raw and loss-postselected fidelity can differ, but as the gate becomes calibrated, the optimal parameter values converge to be the same. We find that this convergence can be a useful proxy for determining whether the gate is optimally calibrated. While in this work we find that our simple calibration method works robustly, more optimal methods could improve the speed of calibration, such as using a more ideal Hessian basis for scanning~\cite{Muniz2025}, and automated calibration via Nelder-Mead optimization or machine learning. \\

\section{Benchmarking protocols}

\begin{figure}
\centering
\includegraphics[width=1\textwidth]{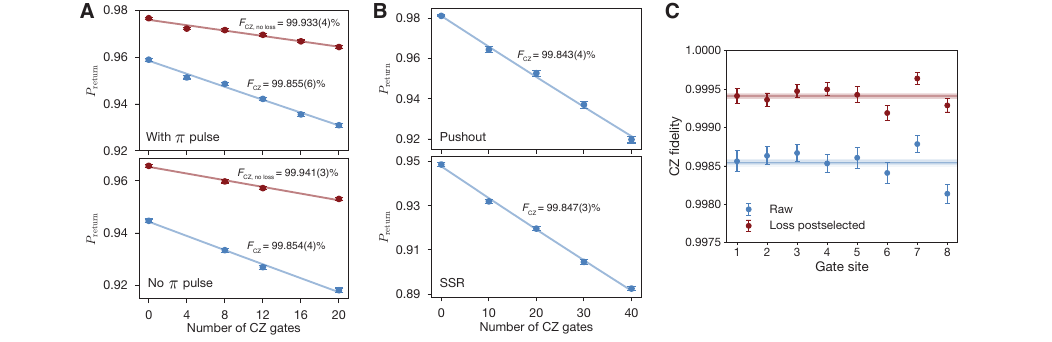}
\caption{\textbf{Readout methods and spatial homogeneity for echo RB.}
Echo RB curves for the data plotted in Fig.~\ref{fig2}C. Measured return probability as a function of number of CZ gates (\textbf{A}) with and without a single-qubit $\pi$ pulse before state-selective readout (\textbf{B}) with pushout versus state-selective readout (SSR). (\textbf{C}) CZ fidelity across the gate sites used for the data in Fig.~\ref{fig2}B.
}
\label{fig:ED_Benchmarking}
\end{figure}

We use global randomized benchmarking (RB) sequences in this work, mostly focusing on RB sequences which include global single-qubit $X$ gates between pairs of CZ gates. We find that this benchmarking scheme is a robust way to calibrate and track the fidelity of CZ gates over time. The results are averaged over 8 CZ gate sites, for which the gate performance is relatively homogeneous (Fig.~\ref{fig:ED_Benchmarking}C). To assess the effects of our readout protocol on the measured fidelity, we perform two variants of this RB sequence: (a) with an additional single-qubit Raman $\pi$ pulse before readout and (b) using resonant blowout of $F=2$ instead of state-selective readout. We observe that all three sequences result in consistent measured fidelities (Fig.~\ref{fig:ED_Benchmarking}A-B).

\begin{figure*}
\centering
\includegraphics[width=1\textwidth]{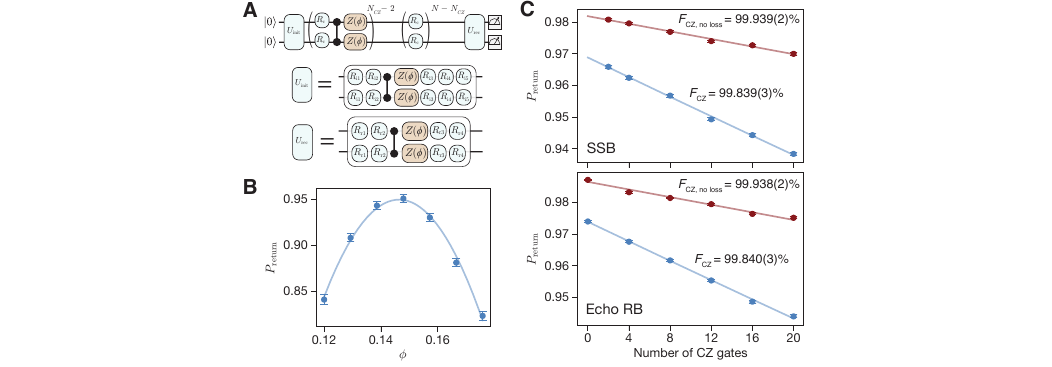}
\caption{\textbf{Symmetric stabilizer benchmarking.}
(\textbf{A}) Circuit used for symmetric stabilizer benchmarking (SSB); figure adapted from Ref.~\cite{Tsai2025}. (\textbf{B}) Optimization of single-particle phase $\phi$ after the CZ gate, here looking at the return probability after 12 CZ gates. (\textbf{C}) Comparison of SSB and echo RB approaches, taken interspersed with each other to avoid any potential bias from experimental drift. Measured fidelities are within error bars of each other.
}
\label{fig:ED_SSB}
\end{figure*}

To check the sensitivity of the measured gate fidelities to the exact benchmarking sequence, we additionally measure the fidelity in the symmetric stabilizer benchmarking (SSB) sequence developed in Ref.~\cite{Tsai2025}. This benchmarking sequence was designed to be insensitive to single-qubit gate errors compared to the echo RB sequence, for which sufficiently large single-qubit error rates can lead to an underestimation of the CZ fidelity~\cite{Tsai2025}. For our experiment, we expect that the single-qubit fidelity is sufficiently high, and that the two benchmarking schemes should offer consistent results \cite{Tsai2025}. In contrast to the echo RB sequence, in which the global $X$ gate between a pair of consecutive CZ gates eliminates the effect of accumulated single-qubit phase during the gate pulse, the SSB sequence is sensitive to the single-qubit phase of the CZ gate, which arises both from the gate trajectory and from the 420-nm differential lightshift on the qubit states. We first calibrate the CZ gate using echo RB, and then optimize the single-qubit phase $Z(\phi)$ in the SSB sequence (Fig.~\ref{fig:ED_SSB}B). We verify that the optimal single-qubit phase is independent of the number of CZ gates, which we find useful for verifying that there is no bias from other possible spurious single-qubit phases in the sequence. We find that the CZ gate fidelity measured with the SSB sequence agrees with that obtained from our echo RB sequence, which shows that the fidelity can be maintained in applications where an echo pulse cannot be applied between adjacent CZ gates.

Finally, we test two different trap configurations for benchmarking the gate. For the CZ gate benchmarking data presented in Figs.~\ref{fig1}-\ref{fig3}, we place two SLM traps close to each other to form the gate site. In quantum circuits with reconfigurable atom positions and connectivity, we typically perform gates between static and moving arrays of traps. In Fig.~\ref{fig4}A, we check that the fidelity is comparable when the gate sites are composed of an AOD and an SLM trap to when both sites are SLM traps. The AOD-SLM trap configuration is used when performing CZ gates for the cluster state and nonlocal circuits presented in Figs.~\ref{fig4}-\ref{fig5}.

\section{CZ gate error sources \label{Sec:error_sources}}

The experimental gate errors can be largely explained by our numerical error model (Fig.~\ref{fig3}B). We compute the errors in the system through a combination of exact diagonalization and simulation of incoherent error sources~\cite{Evered2023,Tsai2025}, with relevant parameters such as $T_1$ and $T_2^*$ informed by experimental measurements. We numerically model the Hamiltonian of our two-photon Rydberg excitation of $^{87}$Rb atoms, which includes the qubit states $|0\rangle, |1\rangle$, three excited states in the 6P$_{3/2}$ hyperfine manifold with $m_F=1$ and $F=1,2,3$, and two Rydberg states with $m_J=+1/2$ and $m_J=-1/2$ (see diagram in Fig.~\ref{fig:ED_ExperimentalSystem}B). We apply the exact smooth-amplitude gate pulse in the numerical model, ignoring the effect of the finite rise time of the AOM. A finite rise time effectively applies a Gaussian window function in the frequency domain to the amplitude profile. The measured rise time of the 420-nm AOMs is around $10$ns, therefore, the smooth pulse profile does not contain high-frequency components which would fall outside the window function.

To model the incoherent error sources, intermediate-state scattering, and Rydberg lifetime, we compare the full density-matrix simulation, including an additional leakage state~\cite{Evered2023}, with calculations using the decay rate equation with the corresponding population in the intermediate states and Rydberg state. We find that the difference between these two methods is negligible. We measure the Rydberg lifetime $T_1$ to be roughly $80\mu$s in our experiment (with Rydberg lasers off), similar to the theoretical prediction of $72\mu$s given by the Alkali Rydberg Calculator \cite{Arc2016}. We assume the lifetime of the intermediate states to be $110$ns \cite{Safronova2011}. To model the effects of intensity, frequency, and position noise of the Rydberg beams, we adapt the Fidelity Response Theory (FRT) developed in Ref.~\cite{Tsai2025} to the two-photon excitation configuration. One key difference for two-photon excitation is that intensity noise can be converted to frequency noise, due to the presence of a large ground-Rydberg lightshift. We numerically find that the relative intensity noise (RIN) and frequency noise of both the $1015$-nm laser, which is filtered by a ULE cavity, and the $420$-nm laser, which is frequency-doubled from a Ti-Sapphire laser~\cite{Denecker2024}, lead to negligible errors. Instead, frequency noise of the gate pulse is dominated by position fluctuations of the laser beams, which leads to time-dependent variations in the lightshifts. This is validated by direct measurements of the $T_2^*$ coherence between $\ket{1}$ and $\ket{r}$ (Table~\ref{tab:experimental_values}). In particular, we measure independent decoherence contributions from finite atomic temperature, compared to frequency noise induced by the $420$-nm and $1015$-nm lasers through three different $\ket{1}$-$\ket{r}$ Ramsey $T_2^*$ measurements. During the wait time between the two $\pi/2$ Rydberg pulses we either (a) keep both Rydberg lasers off, (b) keep only the 1015-nm laser on, or (c) keep only the 420-nm laser on. From the $T_2^*$ measured in the absence of any Rydberg lasers, we estimate the atom temperature to be around $12\mu$K from a Doppler decoherence model, similar to estimates from drop-recapture measurements~\cite{Tuchendler2008}.

We additionally model the position errors of the atoms, which are composed of the motional spread of atoms at finite temperatures and variation in tweezer trap positions. The latter may originate from misalignment or relative focus between the two tweezers, in particular for the AOD-SLM trap configuration, which leads to a variation of the interatomic spacing. As a result, the blockade energy, which scales approximately as $r^{-6}$, will vary across the array, for example causing different detunings due to the light shift from the doubly-occupied Rydberg pair state $\ket{rr}$. At $12\mu$K with a radial trapping frequency of around $80$kHz and axial trapping frequency of around $20$kHz, the atom motional spread has a standard deviation of approximately $70$nm in the radial direction and $270$nm in the axial direction. To estimate the infidelity due to position error in the SLM-SLM trap configuration, where the position spread mainly comes from finite temperature effects, we sample from a Gaussian distribution with a standard deviation of $100$nm ($400$nm) to generate radial (axial) position errors. We obtain the blockade energy $V_{\mathrm{blockade}}(r)$ as a function of gate site spacing $r$ using the Alkali Rydberg Calculator \cite{Arc2016}. We generate $10,000$ sampled blockade energy values and use them in the gate pulse Hamiltonian optimized at $V_{\mathrm{blockade}}=$950 MHz to obtain a mean infidelity of around $0.001\%$ resulting from position errors. In the AOD-SLM configuration, we scan the focal plane of the SLM tweezers and ensure the traps are aligned in the axial direction. We further fine tune the relative in-plane positions of the AOD-SLM tweezers by optimizing on site-resolved fidelities.

\begin{table}
\centering
\begin{tabular}{|l|c|}
\hline
\textbf{Parameter} & \textbf{Value} \\
\hline
Peak two-photon Rabi frequency & $\Omega = 2\pi\times17$ MHz \\
Single-photon 420-nm Rabi frequency & $\Omega_{\text{420}} = 2\pi\times670$ MHz \\
Single-photon 1015-nm Rabi frequency & $\Omega_{\text{1015}} =2\pi\times460$ MHz \\
Intermediate state detuning & $\Delta = 2\pi\times7.8$ GHz \\
$T_2^*$ for $\ket{1}$-$\ket{r}$ with no Rydberg light & $4.7$\,$\mu$s \\
$T_2^*$ for $\ket{1}$-$\ket{r}$ with 420-nm light on & $3.8$\,$\mu$s \\
$T_2^*$ for $\ket{1}$-$\ket{r}$ with 1015-nm light on & $2.8$\,$\mu$s \\
$T_1$ of Rydberg state ($53S$) & 72\,$\mu$s \\
\hline
\end{tabular}
\caption{\label{tab:experimental_values}\textbf{Experiment parameters.} Parameters which are used for our numerical simulations of the experiment, informed by our measurements. The value for $T_1$ is given by the Alkali Rydberg Calculator (ARC)~\cite{Arc2016}.}
\end{table}

\begin{table}[h]
\centering
\label{simulated_fidelity}
\begin{tabular}{|l|c|}
\hline
\textbf{Error source} & \textbf{Smooth-amplitude gate in this work} \\
\hline
Scattering $\ket{1}$ & 0.028\% \\
Scattering $\ket{0}$ & 0.018\% \\
Rydberg $T_1$ & 0.046\% \\
1015-nm $T_2^\ast$ & 0.008\% \\
420-nm $T_2^\ast$ & 0.001\% \\
Doppler $T_2^\ast$ & 0.007\% \\
Rydberg blockade & 0.001\% \\
Rydberg $m_J{\scriptstyle =\frac{1}{2}}$ & 0.002--0.030\% \\
\hline
\textbf{Total fidelity} & \textbf{99.861--99.889\%} \\
\hline
\end{tabular}
\caption{\textbf{Error budget for the smooth-amplitude gate used in this work.} Values plotted in Fig.~\ref{fig3}B for the current configuration. Scattering errors for the qubit state $\ket{0}$ originate from off-resonant scattering from the 420-nm light. The range for the contribution from coupling to the other Rydberg $m_J$ level $\ket{r'}$ relates to optimization of the gate profile (see text).}
\end{table}

We also estimate the error due to off-resonant excitation to the $m_J=+1/2$ Rydberg state $|r'\rangle$. This coupling grows rapidly as we increase the Rabi frequency (Fig.~\ref{ed_fig_fidelity_projections}a,c). This can be alleviated by increasing the Zeeman splitting between the Rydberg states and fine tuning the phase profile of the gate. With $38.5$ MHz Zeeman splitting and $2\pi\times17$ MHz peak Rabi frequency, we numerically find that the error due to coupling to $|r'\rangle$ can be reduced to below $0.005\%$ with further optimization of the phase profile. In particular, the off-resonant excitation fraction is not simply given by $\Omega^2/4\Delta_{\text{Zeeman}}^2$ but can form a nearly closed trajectory with appropriately chosen phase profiles. In Table~\ref{simulated_fidelity}, we specify a range for the $m_J = +1/2$ error, corresponding to how well-optimized the pulse profile is.

\subsection{Understanding loss and leakage errors}

Loss errors in Rydberg gates can arise from several different sources, including Rydberg decay, intermediate state scattering, miscalibration, $T_2^*$, and coupling to the other Rydberg level $\ket{r'}$. Additionally, decay from the Rydberg state to lower-lying excited states and off-resonant scattering from the intermediate state can potentially result in leakage errors into other hyperfine $m_F$ states in the ground-state hyperfine manifold. To estimate the breakdown of different error types, we consider a simple numerical model described in Ref.~\cite{Evered2023}. We assume that any remaining population in the Rydberg state, which can arise from error in the closure of the gate, blackbody-mediated decay to a nearby Rydberg state, or coupling to the other Rydberg state $|r'\rangle$, is converted to loss. This loss conversion occurs because the tweezer traps are turned on within $100$ns after the end of the gate pulse, and the Rydberg state is anti-trapped. With this numerical model, for the parameters used in our experiments, we find that the total numerical error of $0.127\%$ can be decomposed into $0.061\%$ loss error, $0.035\%$ potential leakage error, $0.024\%$ Pauli-Z error, and $0.003\%$ Pauli-X(Y) error. We note that this error budget may overestimate the leakage error and more precise simulations would need to account for several additional effects. For example, excited states are anti-trapped which can lead to loss when the traps are turned back on depending on the exact timing, excited state lifetime, and branching ratios. Additionally, leakage population can be converted to loss by re-excitation in subsequent CZ gates in the benchmarking circuit. Below, we present experimental characterizations which can guide these more detailed numerical simulations.

\begin{figure}
\centering
\includegraphics[width=0.45\textwidth]{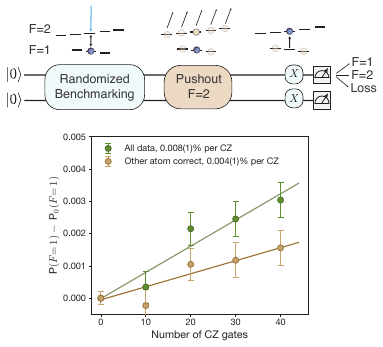}
\caption{\textbf{Understanding leakage in CZ gates.} Measurement of the buildup of population in $F=1$ during the echo RB sequence. At the end of the RB sequence, population in $F=2$ is expelled from the trap with near-resonant pushout light. Then, a single-qubit $\pi$ pulse transfers $\ket{0}$ to $F=2$, and we perform state-selective readout. Plotted is the residual population in $F=1$ as a function of the number of CZ gates, consisting of buildup of leakage population. The green curve is for the raw data, and the tan curve is for postselecting on the other atom being in the correct state at the end of the sequence. 
}
\label{fig:ED_Leakage}
\end{figure}

To understand the effect of leakage into other $m_F$ sublevels, we performed several different measurements. First, as described previously, we compare our randomized benchmarking results with and without the addition of a single-qubit $\pi$ pulse on both qubits before readout (Fig.~\ref{fig2}C and Fig.~\ref{fig:ED_Benchmarking}A). If the leakage errors are strongly asymmetric between $F=2$ and $F=1$ sublevels, we would measure a different CZ gate fidelity for these two protocols. We instead find that the outcomes are consistent with each other.

Second, in Fig.~\ref{fig:ED_Leakage}, we measure leakage into $F=1$ sublevels from the CZ gates~\cite{Tsai2025}. For this, we perform the echo RB sequence with up to 40 CZ gates, and afterwards push out $F=2$ and perform a single-qubit $\pi$ pulse to transfer $F=1, m_F=0$ to $F=2, m_F=0$. Then we use our state-selective readout to separately read out $F=1$ and $F=2$ populations, in which $F=2$ population is optically pumped to the dark state of the state-selective lattice and $F=1$ population remains in a bright state~\cite{Bluvstein2025}. We plot the mean population measured in $F=1$, which consists of population left in the leakage states $m_F=+1$ and $m_F=-1$. We find that the population increases as $0.008(1)\%$ per qubit per CZ. However, often a leakage error on one atom results in a different type of error on the other atom and therefore is detected in our randomized benchmarking. To understand this, we postselect on the other atom in the pair having correctly returned to the $|0\rangle$ state at the end of the sequence, and find that the leakage population is $0.004(1)\%$ per qubit per CZ. This suggests that our measured fidelity might be overestimated by $\sim2\times0.004\%$.
Future work can employ similar techniques combined with population transfer between $m_F$ levels to map out the full leakage populations, in particular for the $F=2$ manifold. 
Additionally, studying how leakage is converted to loss through re-excitation to Rydberg states in subsequent CZ gates is another important avenue for understanding leakage and loss in this high-fidelity regime and their impact in fault-tolerant quantum circuits. Practically, we observe no signature of large-scale correlated loss events down to  the $\sim10^{-5}-10^{-6}$ level in the cluster state and nonlocal circuits (Fig.~\ref{fig:ED_CircuitImplementation}), suggesting that leakage does not lead to detrimental correlated loss events in our complex quantum circuits. 

We additionally consider possible leakage from the 420-nm light driving $m_F$-changing Raman transitions. Polarization impurity alone is insufficient to cause substantial leakage, since a mixture of $\sigma^+$ and $\sigma^-$ light cannot drive $m_F$ transitions in the limit of being far-detuned from $6P_{3/2}$~\cite{Levine2021}. By contrast, if the external magnetic field is at an angle with respect to the beam's propagation axis, then the 420-nm light has a $\pi$ polarization component which can drive transitions with $\Delta m_F=\pm1$. Our choice of $\sigma^-$ polarization results in a vector lightshift that increases the Zeeman splitting between the $m_F$ levels, which is favorable for reducing this effect. We experimentally observe coherent oscillations between $m_F$ levels if we intentionally misalign the magnetic field, consistent with numerical simulations. However, when we carefully align the magnetic field, this effect becomes negligible and does not factor into our error budget.

\begin{figure}
\centering
\includegraphics[width=0.9\textwidth]{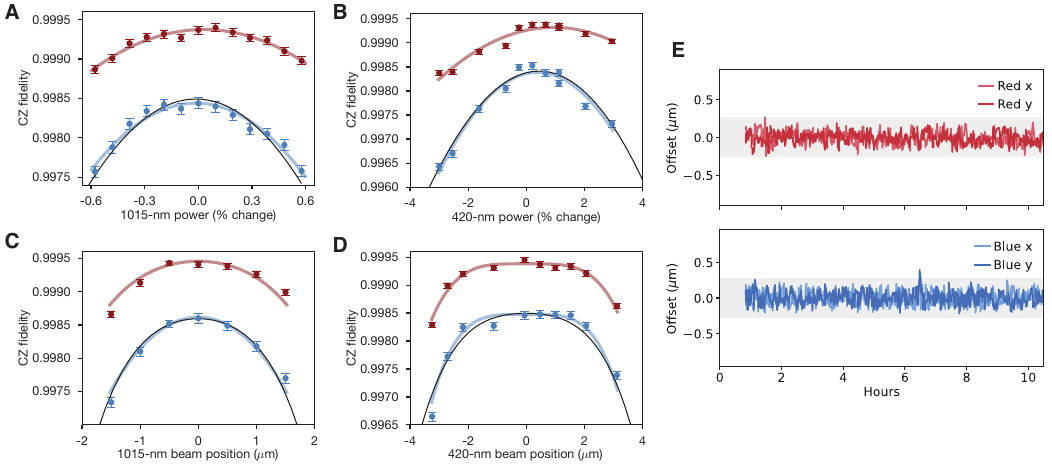}
\caption{\textbf{Understanding sensitivity of CZ fidelity to drifts.}
Sensitivity of the measured CZ gate fidelity to Rydberg beam (\textbf{A-B}) powers and (\textbf{C-D}) positions. Red points are loss-postselected, and blue points are raw CZ fidelities. Faint colored curves are quadratic fits for beam powers and quartic fits for beam positions, and the black curves are from numerical simulations with offsets added as the only free parameter (see text for more details). (\textbf{E}) Positions of Rydberg beams on camera used for tracking and active feedback, averaged between each experimental repetition. These positions were tracked during the data taking for Fig.~\ref{fig2}D.
}
\label{fig:ED_Drifts}
\end{figure}

\subsection{Understanding sensitivity to experimental drift}

In Fig.~\ref{fig2}, we demonstrate that we can robustly achieve roughly 99.85\% fidelity gates for 10 hours without recalibration of the gate. To understand the requirements needed to achieve such stability, we perturb the system and measure how much the fidelity decreases. In particular, for our CZ gates, the main sources of drift are the positions and powers of the Rydberg beams; we find that performance can generally be recovered by tuning the position back and adjusting the powers slightly. For the 1015-nm Rydberg beam, due to the large (roughly 50 MHz) differential light shift on the $\ket{1}-\ket{r}$ transition, the stability requirements are at roughly 0.1\% for intensity and a few hundred nanometers for beam position for our choice of Rydberg parameters and beam size (Fig.~\ref{fig:ED_Drifts}). Since the 420-nm beam is lower in intensity, the gate is less sensitive to perturbations for that beam.

The experimentally measured change in fidelity can be well captured by our numerical modeling with no free parameters other than offsets. We observe that the fidelity change as a function of laser power can be described by a quadratic fit. The sensitivity to the $420$-nm laser intensity is approximately a factor of four weaker than the sensitivity to the $1015$-nm laser intensity, which is consistent with the fact that the lightshift from the $420$-nm beam is about four times smaller than the lightshift from the $1015$-nm beam. This suggests that the primary mechanism for such fidelity reduction is the detuning change from the lightshift. Next, we consider the effect of changing the Rydberg beam positions (Fig.~\ref{fig:ED_Drifts}C-D). For a Gaussian laser beam profile, at the center of the beam where the atoms are located, the intensity changes quadratically as a function of beam displacement, which would result in a quartic response in CZ fidelity to beam position when the detuning is suboptimal. However, the fluctuations of lightshift due to position noise also get worse once atoms are not at the center of the beams and, therefore, the resulting fidelity change can deviate from a quartic fit. Our numerical simulation takes into account the deteriorating $T_2^*$ due to increased sensitivity to pointing noise when the beam is misaligned, and agrees with the experimental data.

\subsection{Choice of $n$}

We choose the same principal quantum number $n=53$ for the Rydberg state as in previous work \cite{Evered2023}. 
Many properties of the Rydberg state depend on $n$, including the Rydberg state lifetime ($\propto n^3$ for radiative decay, $\propto n^2$ for black-body decay), the dipole matrix element of the $1015$-nm transition ($\propto n^{-3/2}$), the blockade strength given by the van der Waals interaction ($\propto n^{11}$), and sensitivity to electric fields ($\propto n^{7}$).
It may seem beneficial to increase $n$ until sensitivity to electric fields starts to limit coherence time $T_2^*$, which reduces errors due to increased Rydberg lifetime and blockade strength. 
However, for two-photon Rydberg excitation, our error model indicates a more stringent tradeoff between coherence time, which is limited by Rydberg beam lightshift fluctuations, and suppression of Rydberg decay. 
The polarizabilities at $1015$nm of the ground state, and of the Rydberg state at our large intermediate state detuning of $7.8$ GHz, remain nearly constant as the transition dipole matrix element changes with $n$. For constant laser intensity, the error contribution from the $1015$-nm lightshift therefore scales as $1/\Omega^2\propto n^3$. In contrast, the error contribution from Rydberg decay scales as $n^{-1/2}$ for black-body decay and as $n^{-3/2}$ for radiative decay. We find that our choice of $n$ is nearly optimal in balancing these different error sources.
In particular, this suggests that the position and intensity stability of the $1015$-nm laser need to be further improved for higher $n$ to be favorable.
This may provide an alternative route to increasing Rabi frequency, in order to suppress Rydberg decay error. However, we additionally note that in circuit contexts, an extra benefit of using lower $n$ is the ability to pack gate sites more densely in the entangling zone, especially within a two-dimensional region, owing to the weaker interaction strength.

\begin{figure}
\centering
\includegraphics[width=\textwidth]{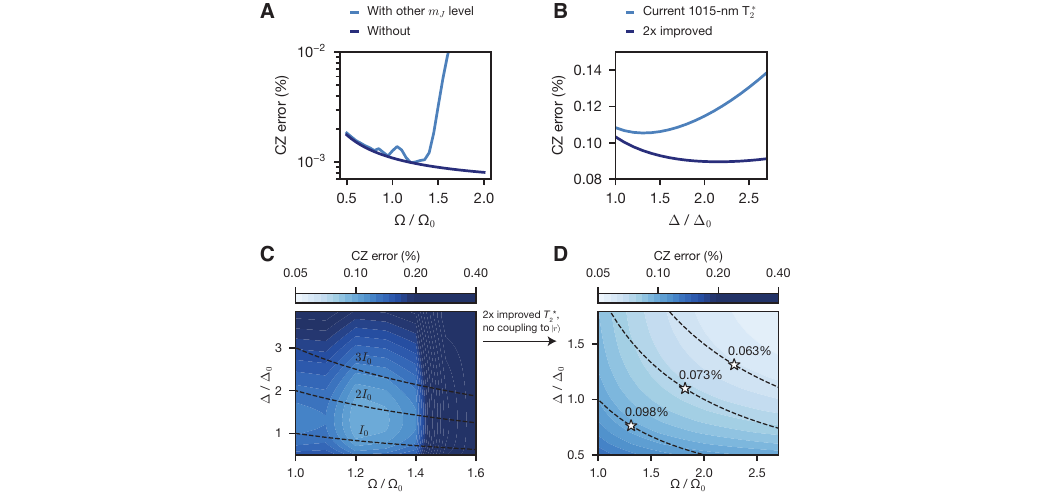}
\caption{\textbf{Path to higher CZ fidelity.} Numerical projection of the CZ fidelity for improved parameters. (\textbf{A}) CZ error as a function of Rabi frequency (relative to current Rabi frequency $\Omega_0$), with and without contributions of the other Rydberg $m_J$ state $\ket{r'}$. (\textbf{B}) CZ error as a function of detuning, relative to the current detuning $\Delta_0$, where the optimum balances contributions from Rydberg $T_2^*$ and intermediate state scattering. The other $m_J$ level is not included for this plot. (\textbf{C}) CZ error as a function of $\Delta$ and $\Omega$, compared to our current Rabi frequency $\Omega_0$ and intermediate state detuning $\Delta_0$ (colorbar is log-scale). Dashed lines denote equal beam intensity, for $1\times$, $2\times$, and $3\times$ current intensity $I_0$ used in this work. The effect of coupling to $\ket{r'}$ prevents going to much higher Rabi frequencies. (\textbf{D}) Projection of CZ fidelity improvements achievable with higher laser intensities, as a function of $\Delta$ and $\Omega$. For this plot, we consider a configuration with $2\times$ improved 1015-nm $T_2^*$ and no coupling to $\ket{r'}$. The white stars mark the optimal configuration that minimizes CZ error for each intensity, with the associated CZ error labeled.}
\label{ed_fig_fidelity_projections}
\end{figure}

\subsection{Path to further improvements in CZ fidelity}

The error budget and our experimental characterization point to a few areas where Rydberg gates could be improved in the future. The error budget is dominated by decay from the Rydberg state due to finite lifetime and by the intermediate excited state in the form of off-resonant scattering. At the current laser intensities, the finite coherence time $T_2^*$ between the qubit state $\ket{1}$ and the Rydberg state $\ket{r}$, which is limited by finite atom temperature and intensity fluctuations of the two Rydberg lasers, also contributes to gate errors. Residual errors mainly come from off-resonant coupling to the other Rydberg level $\ket{r'}$ with $m_J=+1/2$.

Using our numerical error budget, we project that a raw CZ gate fidelity at the level of 99.9-99.95\% should be achievable with this gate scheme. It may seem straightforward to reduce the dominant errors with higher laser intensity. In particular, a higher Rabi frequency reduces the duration of the gate pulse, thereby decreasing error from Rydberg state lifetime. Additionally, a larger intermediate state detuning suppresses off-resonant scattering error during the gate, for which higher laser intensity is necessary to maintain the same Rabi frequency. However, in practice, there are additional experimental challenges, as illustrated in Fig.~\ref{ed_fig_fidelity_projections}C. As the Rabi frequency is increased, we begin to couple strongly to the other Rydberg state $|r'\rangle$ which is shifted up in energy by the bias magnetic field (Fig.~\ref{ed_fig_fidelity_projections}A,C). This off-resonant coupling constitutes a hard limit on our ability to increase Rabi frequency. However, this effect can be mitigated in the future, for example through the use of a higher magnetic field to increase the Zeeman splitting or by coupling through the $6P_{1/2}$ intermediate state where selection rules only allow coupling to one $m_J$ level.

As the intermediate state detuning is increased, the $T_2^*$ error eventually increases considerably due to the high $1015$-nm laser intensities causing large lightshifts on the $\ket{1}-\ket{r}$ transition. The contribution from finite coherence time scales as $(\Omega T_2^*)^{-2}$ or $(\delta f/\Omega)^2$, where $\Omega$ is the two-photon Rabi frequency and $\delta f$ is the integrated frequency noise. In the case where $\delta f$ is dominated by the $420$-nm lightshift fluctuation, it is proportional to the two-photon Rabi frequency $\delta f = \Omega_{420}^2/4\Delta=\Omega/2$ and therefore the infidelity contribution is also a constant, assuming the single-photon Rabi frequencies are balanced between the $420$-nm and $1015$-nm lasers. In contrast, the polarizability from the $1015$-nm laser is nearly a constant when intermediate state detuning is increased away from $7.8$ GHz, for both the ground state and the Rydberg state. As a result, the infidelity due to the $1015$-nm lightshift scales as $(\Omega_{1015}^2/\Omega)^2=\Delta^{2}$, growing quadratically with increasing intermediate state detuning. As a result, by improving the 1015-nm $T_2^*$ by a factor of 2 from our current value, we can increase $\Delta$ further (Fig.~\ref{ed_fig_fidelity_projections}B).

With these two technical improvements of suppressed coupling to $\ket{r'}$ and improved 1015-nm $T_2^*$, the CZ gate fidelity can then be improved further through scaling the laser intensities (Fig.~\ref{ed_fig_fidelity_projections}D). In particular, for a given intensity available for the two Rydberg lasers, there exists an optimal configuration of Rabi frequency and intermediate state detuning which minimizes the error, by making most efficient use of the available laser power. As shown in Fig.~\ref{ed_fig_fidelity_projections}D, this gives a path towards reaching raw CZ gate fidelities approaching the $\sim99.95\%$ level through an increase in available laser power. Fig.~\ref{fig3}B summarizes how the error breakdown changes with these three improvements: better $T_2^*$, suppression of the effect of the other Rydberg level $\ket{r'}$, and higher power excitation.

\begin{figure}
\centering
\includegraphics[width=0.6\textwidth]{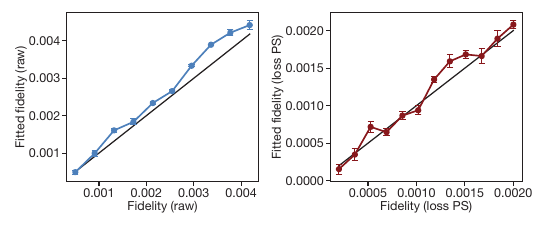}
\caption{\textbf{Numerical simulation of fidelity in cluster state circuit.} Comparison between fitted fidelity of the decay in global return probability and the true CZ fidelity put into the numerical simulation, for the cluster state circuit in Fig.~\ref{fig4}. Black lines denote $y=x$.
}
\label{fig:ED_mirror_fidelity}
\end{figure}

\section{High-fidelity quantum circuits}

To benchmark our ability to perform quantum circuits based on atom movement, we perform the sequence in Fig.~\ref{fig4}B, in which one-dimensional cluster states are created and then unmade repeatedly over time. We then calculate the global return probability, which is defined as the probability that all 20 atoms are measured as $\ket{0}$ after the final $\pi/2$ single-qubit gate used to measure in the $X$ basis. We fit an exponential decay of this global return as a function of the total number of CZ gates applied across the array. For all numbers of CZ gates, we include the same full single-qubit sequence, including all atom moves, in order to isolate the effect of the CZ gates themselves (similar to the RB sequences). In addition to the CZ gates performed between pairs of atoms during the circuit, the unpaired atoms at the edges of the array are also excited to the Rydberg state and can experience gate errors. To be more conservative in our analysis, we do not count these unpaired atoms in the sum for the total number of CZ gates, which leads to a small decrease in the extracted CZ fidelity. Finally, we note that while such a sequence is not as rigorous of a CZ gate benchmarking protocol as the RB sequences, we find in numerical simulations that the extracted fidelity is similar to the CZ fidelity (Fig.~\ref{fig:ED_mirror_fidelity}). Therefore, our measurements suggest that maintaining low-error regimes during quantum circuits is achievable.

\begin{figure}
\centering
\includegraphics[width=0.65\textwidth]{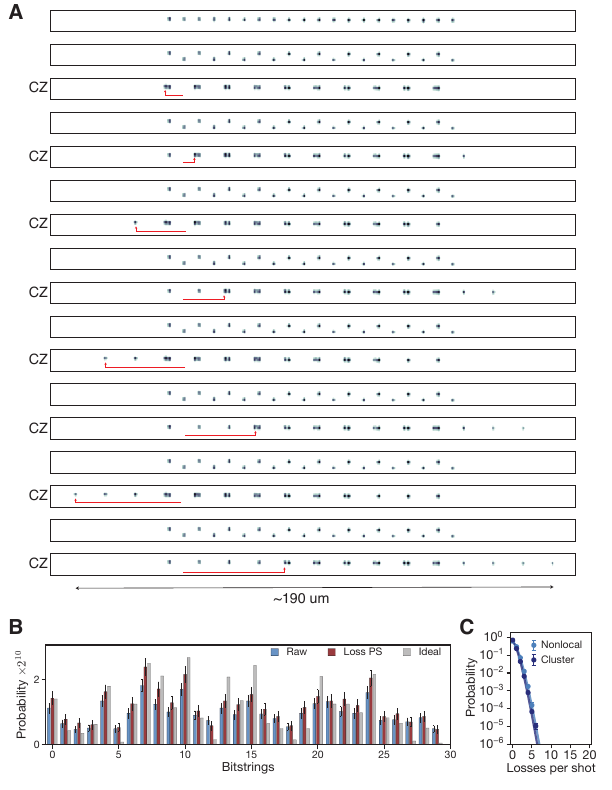}
\caption{\textbf{Nonlocal circuit setup and supplementary data.} (\textbf{A}) Atom images for the CZ gate locations for the nonlocal circuit in Fig.~\ref{fig4} (cluster state circuit repeats the first two nearest-neighbor moves). (\textbf{B}) Sampling from 10-qubit circuit, showing first 30 unordered outcomes.  (\textbf{C}) Distribution of atoms lost per shot (out of 20 atoms total), for the 670,890 shots for the cluster state circuit and the 26,140 shots for the nonlocal circuit. The solid curves are a sum of Poissonian fits of the loss distributions for each different number of CZ gates.
}
\label{fig:ED_CircuitImplementation}
\end{figure}

Fig.~\ref{fig:ED_CircuitImplementation}A shows images of atoms in the locations of the CZ gates used in the nonlocal circuits; for the cluster state circuit, only nearest-neighbor moves are used. The circuit structure is designed such that one half of the atoms are in static SLM traps and the other half are in moving AOD traps, and there are no transfers between different traps during the circuit.

For the nonlocal circuit implementation in Fig.~\ref{fig5}, we find that the measured cross-entropy benchmarking (XEB) results match well to an error budget consisting of a CZ gate Pauli error of $0.06\%$ with $0.09\%$ loss error, initialization Pauli error of $0.36\%$ with $0.37\%$ loss, measurement Pauli error of $0.5\%$ with $0.69\%$ loss, and total error from idling, single-qubit gates, and atom moves during each move layer of $0.091\%$ Pauli error and $0.033\%$ loss error. We further empirically measure the loss during the circuit and upper bound the loss per qubit per move at $<0.03\%$. For the loss per atom per CZ gate measurement shown in Fig.~\ref{fig5}C, we only consider atoms which perform CZ gates in all of the circuit layers, excluding the atoms on the edge of the array. Finally, we find that for both the cluster state and nonlocal circuits, there is no evidence for large-scale correlated loss events, as shown in Fig.~\ref{fig:ED_CircuitImplementation}C.

\section{Theoretical analysis of nonlocal circuits}

\begin{figure}
\centering
\includegraphics[width=\textwidth]{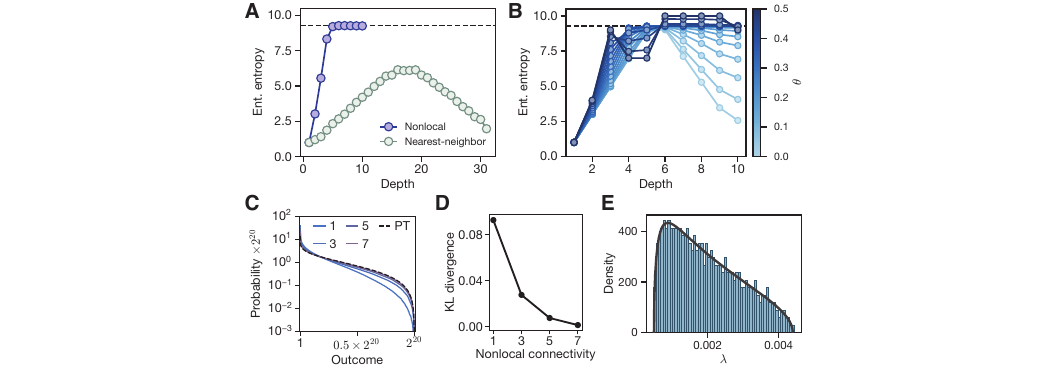}
\caption{\textbf{Theoretical analysis of nonlocal circuits.}
(\textbf{A}) Entanglement entropy as a function of circuit depth, comparing our nonlocal circuit to applying only nearest-neighbor gates at all depths. (\textbf{B}) Entanglement entropy as a function of depth, for different rotation angles $\theta$ for the single-qubit gates in the circuit, where we use $\theta=0.25$ corresponding to a 45-degree $X$ rotation. (\textbf{C}) Distribution of ordered outcome probabilities as a function of nonlocal connectivity, compared to the Porter-Thomas distribution (PT). (\textbf{D}) KL divergence from PT($x$) as a function of nonlocal connectivity. (\textbf{E}) Distribution of eigenvalues of the reduced density matrices for subsystems on the boundary, compared to the Marchenko-Pastur distribution (black curve) expected for Haar-random pure states, for nonlocal connectivity of 11.
}
\label{fig:ED_nonlocal_theory}
\end{figure}

Here we discuss the theoretical properties of our nonlocal circuits described in Fig.~\ref{fig5}. We first numerically study the circuits by evaluating the entanglement entropy, which we define throughout this work as the half-system-size von Neumann entropy (first Renyi entropy), $S_1(A) = -\mathrm{Tr}\left(\rho_A \log \rho_A\right)$, where we consider the half system $A$ with a cut in the middle of the chain. Specifically, we choose nonlocal quantum circuits that generate states with entanglement entropy close to the Page entropy \cite{PhysRevLett.71.1291}, which has a value of $\sim 9.28$ for 20 qubits. We find that the circuit structure shown in Fig.~\ref{fig5}A gives rise to super-ballistic scrambling, in which the time to saturate the Page entropy scales as $\sqrt{N}$ for system size $N$. The intuition underlying this scaling is that the range of CZ gate connectivity expands linearly with circuit depth, as compared to the case of only nearest-neighbor CZ gates where the connectivity is constant with depth. Further, performing the same circuit with only nearest-neighbor gates causes the entanglement entropy not to saturate at the Page entropy (Fig.~\ref{fig:ED_nonlocal_theory}A). In this case, the failure to reach the Page entropy is attributed to the integrability of the circuit dynamics, which constrains the system’s evolution and leads to revivals in the block entanglement entropy \cite{Modak_2020}. Note that a modification of the nearest-neighbor circuit, such as additional single-qubit $z$-axis rotations in the bulk $e^{-i \phi \sum_j Z_j}$, could break integrability and lead to scrambling, but in a time that scales linearly with system size $N$.
An additional component of our circuits is the choice of the 45-degree single-qubit rotation $R = X(\pi/4)$ between CZ layers. We explore this rotation angle in theory, finding that larger or smaller angles do not saturate to the Page theory value (Fig.~\ref{fig:ED_nonlocal_theory}B). 

When sampling the output of these circuits in the Pauli $Z$ basis, we find that the resulting ideal distribution of outcome probabilities approaches the Porter-Thomas distribution PT($x$) as a function of depth (Fig.~\ref{fig:ED_nonlocal_theory}C). We can compare between our circuit's output distribution q($x$) and PT($x$) more quantitatively by calculating the KL divergence, $D_{\text{KL}} = \sum_{x} \text{PT}(x)\,\log\frac{\text{PT}(x)}{q(x)}$. In Fig.~\ref{fig:ED_nonlocal_theory}D, we find that $D_{\text{KL}}$ approaches zero for our circuits with sufficient depth. Finally, we characterize the scrambling in these nonlocal circuits more rigorously by examining the distribution of eigenvalues of the reduced density matrices for subsystems on the boundary. In particular, for sufficiently deep circuits (nonlocal connectivity of 11), we find that this eigenvalue distribution is well-described by a Marchenko–Pastur distribution \cite{Collins_2010}, consistent with the entanglement spectrum expected for Haar-random pure states (Fig.~\ref{fig:ED_nonlocal_theory}E).

\end{document}